# Multivariate vector sampling expansions in shift invariant subspaces<sup>☆</sup>

Qingyue Zhang

*Department of Mathematics and LPMC, Nankai University, Tianjin 300071, China*

**Abstract**

In this paper, we study multivariate vector sampling expansions on general finitely generated shift-invariant subspaces. Necessary and sufficient conditions for a multivariate vector sampling theorem to hold are given.

## 1. Introduction and Main Results

If $H$ is a Hilbert space, we define $H^{(q)} = H \times H \times \cdots \times H(q \text{ term})$. Given $f = (f_1, f_2, \cdots, f_q)^T, g = (g_1, g_2, \cdots, g_q)^T \in H^{(q)}$, the inner product $f$ and $g$ is defined by

$$\langle f, g \rangle_{H^{(q)}} = \sum_{p=1}^{q} \langle f_p, g_p \rangle_H.$$

Let $\varphi_j = (\varphi_{j,1}, \varphi_{j,2}, \cdots, \varphi_{j,r})^T \in L^2(\mathbb{R}^d)^{(r)}, 1 \leq j \leq N$ be a stable generator for the shift-invariant subspace

$$V_\varphi^2 := \left\{ \sum_{j=1}^{N} \sum_{\alpha \in \mathbb{Z}^d} a_{j,\alpha} \varphi_j(\cdot - \alpha) : \left\{ a_{j,\alpha} : 1 \leq j \leq N, \alpha \in \mathbb{Z}^d \right\} \in \ell^2(\mathbb{Z}^d)^{(N)} \right\}.$$

i.e., the sequence $\{\varphi_j(\cdot - \alpha) : 1 \leq j \leq N, \alpha \in \mathbb{Z}^d\}$ is a Riesz basis for $V_\varphi^2$. Recall that $\{\varphi_j(\cdot - \alpha) : 1 \leq j \leq N, \alpha \in \mathbb{Z}^d\}$ is a Riesz basis for $V_\varphi^2$, if there

---

<sup>☆</sup>This work was supported partially by the National Natural Science Foundation of China (10971105 and 10990012) and the Natural Science Foundation of Tianjin (09JCY-BJC01000).

*Email address:* `jczhangqingyue@mail.nankai.edu.cn` (Qingyue Zhang)

exist two constants $A, B > 0$ such that for any $\{c_{j,\alpha} : 1 \leq j \leq N, \alpha \in \mathbb{Z}^d\} \in \ell^2(\mathbb{Z}^d)^{(N)}$ one has

$$A \sum_{j=1}^{N} \sum_{\alpha \in \mathbb{Z}^d} |c_{j,\alpha}|^2 \leq \left\| \sum_{j=1}^{N} \sum_{\alpha \in \mathbb{Z}^d} c_{j,\alpha} \varphi_j(\cdot - \alpha) \right\|_{L^2(\mathbb{R}^d)^{(r)}}^2 \leq B \sum_{j=1}^{N} \sum_{\alpha \in \mathbb{Z}^d} |c_{j,\alpha}|^2.$$

We assume throughout the paper that the vector functions in the shift-invariant subspace $V_\varphi^2$ are continuous on $\mathbb{R}^d$. Equivalently(see [1] or [2]), that the generator $\varphi_j, 1 \leq j \leq N$ is continuous on $\mathbb{R}^d$ and

$$\sup_{x \in \mathbb{R}^d} \sum_{j=1}^{N} \sum_{p=1}^{r} \sum_{\alpha \in \mathbb{Z}^d} |\varphi_{j,p}(x - \alpha)|^2 < \infty.$$

If $V_{\varphi_j}^2 (1 \leq j \leq N)$ is defined by

$$V_{\varphi_j}^2 := \left\{ \sum_{\alpha \in \mathbb{Z}^d} a_\alpha \varphi_j(\cdot - \alpha) : \{a_\alpha : \alpha \in \mathbb{Z}^d\} \in \ell^2(\mathbb{Z}^d) \right\}.$$

Then we have

$$V_\varphi^2 = \sum_{j=1}^{N} V_{\varphi_j}^2.$$

Define $T_{\varphi_j} : L^2\left([(j-1)/N, j/N)^d\right) \longrightarrow V_{\varphi_j}^2$, by

$$T_{\varphi_j} F = \sum_{\alpha \in \mathbb{Z}^d} c_{F,j,\alpha} \varphi_j(\cdot - \alpha),$$

where $c_{F,j,\alpha} = N^{d/2} \int_{[(j-1)/N, j/N)^d} F(x) e^{2\pi N i \alpha^T \cdot x} dx$.

For convenience, in this paper we make $\chi_p(x) = \chi_{[(p-1)/N, p/N)^d}(x)$ and $e_\alpha(x) = N^{d/2} e^{-2\pi N i \alpha^T \cdot x}$.

**Lemma 1.1.** $T_\varphi = \sum_{j=1}^{N} T_{\varphi_j}$ is an isomorphism between $L^2[0,1)^d$ and $V_\varphi^2$.

**Proof**. Since the sequence $\{\varphi_j(\cdot - \alpha) : 1 \leq j \leq N, \alpha \in \mathbb{Z}^d\}$ is a Riesz basis for $V_\varphi^2$, Therefore, for any $F \in L^2[0,1)^d$, we have

$$\|T_\varphi F\|_{L^2(\mathbb{R}^d)^{(r)}}^2 \leq B \sum_{j=1}^{N} \sum_{\alpha \in \mathbb{Z}^d} |c_{F,j,\alpha}|^2 = B \sum_{j=1}^{N} \left\| F \chi_{[(j-1)/N, j/N]^d} \right\|_{L^2[0,1)^d}^2$$

$$= B \|F\|_{L^2[0,1)^d}^2.$$

Similarly, we also have
$$\|T_\varphi F\|^2_{L^2(\mathbb{R}^d)^{(r)}} \geq A\|F\|^2_{L^2[0,1)^d}.$$
□

Given a nonsingular matrix $M$ with integer entries. Let $\gamma_k + M^T\mathbb{Z}^d, 1 \leq k \leq m = |\det(M)|$ be the $m$ distinct elements of the coset space $\mathbb{Z}^d/M^T\mathbb{Z}^d$ with $\gamma_1 = 0$. Define $Q_k = M^{-T}\gamma_k/N + M^{-T}[0,1)^d/N, 1 \leq k \leq m$, we have (see [6, P.110])

$$Q_k \cap Q_{k'} = \emptyset \text{ and } \mathrm{Vol}\left(\bigcup_{k=1}^m Q_k\right) = \frac{1}{N}.$$

Thus, for any function $F$ integrable in $[0, 1/N)^d$ and $\mathbb{Z}^d/N$-periodic, we have $\int_{[0,1/N)^d} F(x)dx = \sum_{k=1}^m \int_{Q_k} F(x)dx$.

Let $g_j \in L^2[0,1)^d, 1 \leq j \leq s$, define

$$G_p(x) := \begin{bmatrix} g_1(x)\chi_p(x) & \cdots & g_1(x)\chi_p(x + M^{-T}\gamma_m/N) \\ g_2(x)\chi_p(x) & \cdots & g_2(x)\chi_p(x + M^{-T}\gamma_m/N) \\ \vdots & \vdots & \vdots \\ g_s(x)\chi_p(x) & \cdots & g_s(x)\chi_p(x + M^{-T}\gamma_m/N) \end{bmatrix}, 1 \leq p \leq N. \quad (1.1)$$

and its related constants

$$A_G := \min_{1 \leq p \leq N} \operatorname*{ess\,inf}_{x \in [0,1/N)^d} \lambda_{min}\left[G_p^*(x)G_p(x)\right],$$

$$B_G := \max_{1 \leq p \leq N} \operatorname*{ess\,sup}_{x \in [0,1/N)^d} \lambda_{max}\left[G_p^*(x)G_p(x)\right].$$

**Lemma 1.2.** *Suppose that $g_j \in L^2[0,1)^d, 1 \leq j \leq s$ and $G_p(x), 1 \leq p \leq N$ is its associated matrix as in (1.1). Then*

(a) *The sequence $\left\{\overline{g_j(x)}\chi_p(x)e_\alpha(M^Tx) : 1 \leq j \leq s, 1 \leq p \leq N, \alpha \in \mathbb{Z}^d\right\}$ is a complete system for $L^2[0,1)^d$ if and only if for any $1 \leq p \leq N$ the rank of the matrix $G_p(x)$ is $m$ a.e. in $[0, 1/N)^d$.*

(b) *The sequence $\left\{\overline{g_j(x)}\chi_p(x)e_\alpha(M^Tx) : 1 \leq j \leq s, 1 \leq p \leq N, \alpha \in \mathbb{Z}^d\right\}$ is a bessel sequence for $L^2[0,1)^d$ if and only if $B_G < \infty$. In this case, the optimal Bessel bound is $B_G/m$.*

(c) *The sequence $\left\{\overline{g_j(x)}\chi_p(x)e_\alpha(M^Tx) : 1 \leq j \leq s, 1 \leq p \leq N, \alpha \in \mathbb{Z}^d\right\}$ is a frame for $L^2[0,1)^d$ if and only if $0 < A_G \leq B_G < \infty$. In this case, the optimal frame bounds is $A_G/m$ and $B_G/m$.*

(d) The sequence $\{\overline{g_j(x)}\chi_p(x)e_\alpha(M^Tx) : 1 \leq j \leq s, 1 \leq p \leq N, \alpha \in \mathbb{Z}^d\}$ is a Riesz basis for $L^2[0,1)^d$ if and only if it is a frame and $s = m$.

**Proof.** For any $F \in L^2[0,1)^d$, we have

$$\left\langle F(x), \overline{g_j(x)}\chi_p(x)e_\alpha(M^Tx)\right\rangle_{L^2[0,1)^d}$$

$$= \int_{[0,1)^d} F(x)g_j(x)\overline{\chi_p(x)}\,\overline{e_\alpha(M^Tx)}dx$$

$$= \int_{[0,1)^d} F(x)g_j(x)\chi_p(x)e^{2\pi Ni\alpha^T\cdot M^Tx}dx$$

$$= \int_{[0,1)^d} F(x)\chi_p(x)g_j(x)\chi_p(x)e^{2\pi Ni\alpha^T\cdot M^Tx}dx$$

$$= \int_{[0,1/N)^d} F(x)\chi_p(x)g_j(x)\chi_p(x)e^{2\pi Ni\alpha^T\cdot M^Tx}dx$$

$$= \sum_{k=1}^{m}\int_{\mathring{Q}_k} F(x)\chi_p(x)g_j(x)\chi_p(x)e^{2\pi Ni\alpha^T\cdot M^Tx}dx$$

$$= \int_{M^{-T}[0,1/N)^d} \sum_{k=1}^{m}(F\chi_p)(x + M^{-T}\gamma_k/N) \times$$
$$(g_j\chi_p)(x + M^{-T}\gamma_k/N)N^{d/2}e^{2\pi Ni\alpha^T\cdot M^Tx}dx. \qquad (1.2)$$

where we have considered the $\mathbb{Z}^d/N$-periodic extension of $F$. Then

$$\sum_{j=1}^{s}\sum_{\alpha\in\mathbb{Z}^d}\left|\left\langle F(x), \overline{g_j(x)}\chi_p(x)e_\alpha(M^Tx)\right\rangle_{L^2[0,1)^d}\right|^2$$

$$= \frac{1}{m}\sum_{j=1}^{s}\left\|\sum_{k=1}^{m}(F\chi_p)(x + M^{-T}\gamma_k/N) \times\right.$$
$$\left.(g_j\chi_p)(x + M^{-T}\gamma_k/N)\right\|^2_{L^2(M^{-T}[0,1/N)^d)}.$$

Denoting $\mathbb{F}_p(x) := \big((F\chi_p)(x), \cdots, (F\chi_p)(x + M^{-T}\gamma_m/N)\big)^T$, the above reads

$$\sum_{j=1}^{s}\sum_{\alpha\in\mathbb{Z}^d}\left|\left\langle F(x), \overline{g_j(x)}\chi_p(x)e_\alpha(M^Tx)\right\rangle_{L^2[0,1)^d}\right|^2$$

$$= \frac{1}{m}\|G_p(x)\mathbb{F}_p(x)\|^2_{L^2(M^{-T}[0,1/N)^d)^{(s)}}. \qquad (1.3)$$

Since $\gamma_1, \gamma_2, \cdots, \gamma_m$ are representatives of distinct cosets of $\mathbb{Z}^d/M^T\mathbb{Z}^d$, therefore for any $1 \leq k \leq m$ the matrix $G(x + M^{-T}\gamma_k/N)$ has the same columns of $G(x)$. Hence, for any $1 \leq p \leq N$ rank $G_p(x) = m$ a.e. in $[0, 1/N)^d$ if and only if rank $G_p(x) = m$ a.e. in $M^{-T}[0, 1/N)^d$. Moreover, we have

$$\operatorname*{ess\,inf}_{x \in [0,1/N)^d} \lambda_{min}\left[G_p^*(x)G_p(x)\right] = \operatorname*{ess\,inf}_{x \in M^{-T}[0,1/N)^d} \lambda_{min}\left[G_p^*(x)G_p(x)\right]$$

and

$$\operatorname*{ess\,sup}_{x \in [0,1/N)^d} \lambda_{min}\left[G_p^*(x)G_p(x)\right] = \operatorname*{ess\,sup}_{x \in M^{-T}[0,1/N)^d} \lambda_{min}\left[G_p^*(x)G_p(x)\right]$$

To prove $(a)$, assume that there exists set $\Omega \subseteq M^{-T}[0, 1/N)^d$ with positive measure and $1 \leq p_0 \leq N$ such that rank $G_{p_0}(x) < m$, $x \in \Omega$. Then, there exists a measurable function $v(x)$, such that $G_{p_0}(x)v(x) = 0$ and $|v(x)| = 1$ in $\Omega$. Define $F \in L^2[0, 1)^d$ such that $\mathbb{F}_{p_0}(x) = v(x)$ if $x \in \Omega$, $\mathbb{F}_{p_0}(x) = 0$ if $x \in M^{-T}[0, 1/N)^d \setminus \Omega$ and $\mathbb{F}_p(x) = 0$ if $p \neq p_0$. Hence, from (1.3) we obtain the system is not complete. conversely, if the system is not complete, by using (1.3) we obtain a $\mathbb{F}_{\bar{p}}(x)$ different from 0 in a set with positive measure such that $G_{\bar{p}}(x)\mathbb{F}_{\bar{p}}(x) = 0$. Thus rank $G_{\bar{p}}(x) < m$ on a set with positive measure.

If $B_G < \infty$ then, for each $F \in L^2[0, 1)^d$, we have

$$\sum_{p=1}^{N}\sum_{j=1}^{s}\sum_{\alpha \in \mathbb{Z}^d} \left|\left\langle F(x), \overline{g_j(x)}\chi_p(x)e_\alpha(M^T x)\right\rangle_{L^2[0,1)^d}\right|^2$$

$$= \sum_{p=1}^{N} \frac{1}{m} \|G_p(x)\mathbb{F}_p(x)\|^2_{L^2\left(M^{-T}[0,1/N)^d\right)^{(s)}}$$

$$\leq B_G \frac{1}{m} \sum_{p=1}^{N} \|\mathbb{F}_p(x)\|^2_{L^2\left(M^{-T}[0,1/N)^d\right)^{(s)}}$$

$$= B_G \frac{1}{m} \sum_{p=1}^{N} \int_{M^{-T}[0,1/N)^d} \sum_{k=1}^{m} \left|(F\chi_p)(x + M^{-T}\gamma_k/N)\right|^2 dx$$

$$= B_G \frac{1}{m} \sum_{p=1}^{N}\sum_{k=1}^{m} \int_{Q_k} |(F\chi_p)(x)|^2 dx$$

$$= B_G \frac{1}{m} \sum_{p=1}^{N} \int_{[0,1/N)^d} |(F\chi_p)(x)|^2 dx$$

$$= B_G \frac{1}{m} \sum_{p=1}^{N} \int_{[(p-1)/N, p/N]^d} |F(x)|^2 \, dx = \frac{B_G}{m} \|F\|^2_{L^2[0,1]^d}$$

Hence $\left\{\overline{g_j(x)}\chi_p(x)e_\alpha(M^T x) : 1 \leq j \leq s, 1 \leq p \leq N, \alpha \in \mathbb{Z}^d\right\}$ is a bessel sequence for $L^2[0,1]^d$ and the optimal Bessel bound is less than or equal to $\frac{B_G}{m}$.

Let $K < B_G$. Then, there a set $\Omega_K \subset M^{-T}[0, 1/N)^d$ with positive measure and $1 \leq \tilde{p} \leq N$ such that $\lambda_{max}\left[G^*_{\tilde{p}}(x)G_{\tilde{p}}(x)\right] \geq K$ for $x \in \Omega_K$. Let $F \in L^2[0,1]^d$ such that its associated vector function $\mathbb{F}_{\tilde{p}}(x)$ is 0 if $x \in M^{-T}[0, 1/N)^d \setminus \Omega_K$, $\mathbb{F}_{\tilde{p}}(x)$ is an eigenvector of norm 1 associated with the largest eigenvalue of $G^*_{\tilde{p}}(x)G_{\tilde{p}}(x)$ if $x \in \Omega_K$ and $\mathbb{F}_p(x) = 0$ if $p \neq \tilde{p}$. Using (1.3), we obtain

$$\sum_{j=1}^{s} \sum_{\alpha \in \mathbb{Z}^d} \left|\left\langle F(x), \overline{g_j(x)}\chi_{\tilde{p}}(x)e_\alpha(M^T x)\right\rangle_{L^2[0,1]^d}\right|^2$$
$$\geq \frac{1}{m} \int_{M^{-T}[0,1/N)^d} K |\mathbb{F}_{\tilde{p}}(x)|^2 \, dx = \frac{K}{m}\|F\|^2_{L^2[0,1]^d}.$$

Therefore if $B_G = \infty$ then $\left\{\overline{g_j(x)}\chi_p(x)e_\alpha(M^T x) : 1 \leq j \leq s, 1 \leq p \leq N, \alpha \in \mathbb{Z}^d\right\}$ is not a bessel sequence for $L^2[0,1]^d$, and if $B_G < \infty$ then the optimal Bessel bound is $B_G/m$. This completes the proof of (b). The proofs of (c) are completely analogous.

To prove (d), we assume that $m = s$ and that the sequence is a frame. We see that it is a Riesz basis by proving that the analysis operator

$$\Lambda : L^2[0,1]^d \to \ell^2(\mathbb{Z}^d)^{(N \times s)},$$

i.e.

$$\Lambda(F) := \left\{\left\langle F(x), \overline{g_j(x)}\chi_p(x)e_\alpha(M^T x)\right\rangle_{L^2[0,1]^d} : \right.$$
$$\left. 1 \leq j \leq s, 1 \leq p \leq N, \alpha \in \mathbb{Z}^d\right\}.$$

is surjective (see [7, Theorem 6.5.1]). To this end, notice that when $m = s$ for any $1 \leq p \leq N$ the matrix $G_p, 1 \leq p \leq N$ is a square matrix and hence, the condition $A_G > 0$ implies that for any $1 \leq p \leq N$ the inverse matrix $G_p^{-1}(x)$ exists and its entries are essentially bounded. For $1 \leq p \leq N$, let $\{c_{j,\alpha}^p : 1 \leq j \leq s, \alpha \in \mathbb{Z}^d\}$ be an element of $\ell^2(\mathbb{Z}^d)^{(s)}$. For $1 \leq p \leq N, 1 \leq j \leq s$ we define the function

$$\xi_j^p(x) := m \sum_{\alpha \in \mathbb{Z}^d} c_{j,\alpha}^p e_\alpha(M^T x),$$

and let $F$ be the function such that

$$\mathbb{F}_p(x) = G_p^{-1}(x)\left(\xi_1^p(x), \xi_2^p(x), \cdots, \xi_s^p(x)\right)^T, x \in M^{-T}[0, 1/N)^d.$$

This function belongs to $L^2[0,1)^d$ because the entries of $G_p^{-1}(x)$ are essentially bounded. We have that $G_p(x)\mathbb{F}_p(x) = (\xi_1^p(x), \xi_2^p(x), \cdots, \xi_s^p(x))^T$, and using (1.2) we obtain that

$$\left\langle F(x), \overline{g_j(x)}\chi_p(x)e_\alpha(M^T x)\right\rangle_{L^2[0,1)^d}$$
$$= \int_{M^{-T}[0,1/N)^d} \xi_j^p(x) N^{d/2} e^{2\pi N i \alpha^T \cdot M^T x} = c_{j,\alpha}^p.$$

and consequently $\Lambda(F) = \left\{c_{j,\alpha}^p : 1 \leq j \leq s, 1 \leq p \leq N, \alpha \in \mathbb{Z}^d\right\}$.

Conversely, if $\left\{\overline{g_j(x)}\chi_p(x)e_\alpha(M^T x) : 1 \leq j \leq s, 1 \leq p \leq N, \alpha \in \mathbb{Z}^d\right\}$ is a Riesz basis. Let $\{f_{j,p,\alpha} : 1 \leq j \leq s, 1 \leq p \leq N, \alpha \in \mathbb{Z}^d\}$ be its dual Riesz basis. Then, by using (1.2) we obtain for $1 \leq p \leq N$

$$m\delta_{\alpha,0}\delta_{j,j'} = \int_{M^{-T}[0,1/N)^d} \sum_{k=1}^m (f_{j',p,0}\chi_p)(x + M^{-T}\gamma_k/N) \times$$
$$(g_j\chi_p)(x + M^{-T}\gamma_k/N) N^{d/2} e^{2\pi N i \alpha^T \cdot M^T x} dx$$

Therefore, for $1 \leq j, j' \leq s$, we have

$$\sum_{k=1}^m (f_{j',p,0}\chi_p)(x + M^{-T}\gamma_k/N)(g_j\chi_p)(x + M^{-T}\gamma_k/N) = m\delta_{j,j'}, a.e.$$

Thus the matrix $G(x)$ has a right inverse; in particular, $s \leq m$. As a consequence (a) we have $s \geq m$ and, finally, $s = m$. $\square$

We consider $s$ linear-invariant systems $\mathcal{L}_j, 1 \leq j \leq s$ in $L^2(\mathbb{R}^d)^{(r)}$ such that for any $f = (f_1, f_2, \cdots, f_r)^T \in L^2(\mathbb{R}^d)^{(r)}$,

$$(\mathcal{L}_j f)(t) = [f * P](t) = \sum_{q=1}^r \int_{\mathbb{R}^d} f_q(x) p_{j,q}(t-x) dx,$$

where $P(x)$ is an $s \times r$ matrix with entries $p_{j,q} \in L^1(\mathbb{R}^d), 1 \leq j \leq s, 1 \leq q \leq r$. Let $h_j(t) = \left(\overline{p_{j,1}(-t)}, \overline{p_{j,2}(-t)}, \cdots, \overline{p_{j,r}(-t)}\right)^T$, we have

$$(\mathcal{L}_j f)(t) = \langle f(\cdot), h_j(\cdot - t) \rangle_{L^2(\mathbb{R}^d)^{(r)}}.$$

The set of systems $\{\mathcal{L}_1, \mathcal{L}_2, \cdots, \mathcal{L}_s\}$ is an $M$-stable filtering sampler for $V_\varphi^2$ if there exist two positive constants $C_1$ and $C_2$ such that [5] for any $f = f^{(1)} + f^{(2)} + \cdots + f^{(N)} \in V_\varphi^2$ where $f^{(p)} \in V_{\varphi_p}^2$, we have

$$C_1 \|f\|_{L^2(\mathbb{R}^d)^{(r)}}^2 \leq \sum_{p=1}^N \sum_{j=1}^s \sum_{\alpha \in \mathbb{Z}^d} \left|\mathcal{L}_j f^{(p)}(M\alpha)\right|^2 \leq C_2 \|f\|_{L^2(\mathbb{R}^d)^{(r)}}^2.$$

For $1 \leq j \leq s$ and $1 \leq p \leq N$, we define $g_{j,p}(x)$ by

$$g_{j,p}(x) := \sum_{\alpha \in \mathbb{Z}^d} (\mathcal{L}_j \varphi_p)(\alpha) e^{-2\pi N i \alpha^T x}. \tag{1.4}$$

**Lemma 1.3.** *Let $f$ be a function in $V_\varphi^2$ such that $f = f^{(1)} + f^{(2)} + \cdots + f^{(N)}$ where $f^{(p)} \in V_{\varphi_p}^2$ and $f^{(p)} = \mathrm{T}_{\varphi_p} F_p, F_p \in L^2\left([(p-1)/N, p/N)^d\right)$. For every $1 \leq j \leq s$, we have*

$$(\mathcal{L}_j f^{(p)})(M\beta) = \left\langle F_p(\cdot), \overline{g_{j,p}(\cdot)} e_\beta(M^T \cdot) \right\rangle_{L^2[0,1)^d}, \quad \beta \in \mathbb{Z}^d. \tag{1.5}$$

**Proof.** For each $\beta \in \mathbb{Z}^d$ we have

$$\begin{aligned}
(\mathcal{L}_j f^{(p)})(M\beta) &= \langle f^{(p)}(\cdot), h_j(\cdot - M\beta) \rangle_{L^2(\mathbb{R}^d)^{(r)}} \\
&= \left\langle \sum_{\alpha \in \mathbb{Z}^d} c_{F_p, \alpha} \varphi_p(\cdot - \alpha), h_j(\cdot - M\beta) \right\rangle_{L^2(\mathbb{R}^d)^{(r)}} \\
&= \sum_{\alpha \in \mathbb{Z}^d} c_{F_p, \alpha} \langle \varphi_p(\cdot - \alpha), h_j(\cdot - M\beta) \rangle_{L^2(\mathbb{R}^d)^{(r)}} \\
&= \sum_{\alpha \in \mathbb{Z}^d} c_{F_p, \alpha} (\mathcal{L}_j \varphi_p)(M\beta - \alpha) \\
&= \sum_{\alpha \in \mathbb{Z}^d} \langle F_p(\cdot), e_\alpha(\cdot) \rangle_{L^2\left([(p-1)/N, p/N)^d\right)} (\mathcal{L}_j \varphi_p)(M\beta - \alpha) \\
&= \left\langle F_p(\cdot), \sum_{\alpha \in \mathbb{Z}^d} \overline{(\mathcal{L}_j \varphi_p)}(M\beta - \alpha) e_\alpha(\cdot) \right\rangle_{L^2\left([(p-1)/N, p/N)^d\right)} \\
&= \left\langle F_p(\cdot), \overline{g_{j,p}(\cdot)} e_\beta(M^T \cdot) \right\rangle_{L^2\left([(p-1)/N, p/N)^d\right)}.
\end{aligned}$$

□

**Theorem 1.4.** *Assume that the function $g_{j,p}(x)$ given in (1.4) belong to $L^\infty\left([(p-1)/N, p/N)^d\right)$ for each $1 \le j \le s$ and $1 \le p \le N$. Let $G_p(x)$ be the associated matrix define in $[(p-1)/N, p/N)^d$ as in (1.1). The following statements are equivalents:*

(a) $A_G > 0$;

(b) *The set of systems* $\{\mathcal{L}_1, \mathcal{L}_2, \cdots, \mathcal{L}_s\}$ *is an $M$-stable filtering sampler for $V_\varphi^2$;*

(c) *For $1 \le p \le N$, there exist vectors $(d_1^p(x), d_2^p(x), \cdots, d_s^p(x))$ with entries $d_j^p \in L^\infty\left([(p-1)/N, p/N)^d\right)$ satisfying*

$$(d_1^p(x), d_2^p(x), \cdots, d_s^p(x))G_p(x) = (1, 0, \cdots, 0)$$
$$a.e. \, in \, [(p-1)/N, p/N)^d; \quad (1.6)$$

(d) *There exists a frame for $V_\varphi^2$ having the form $\{S_j^p(t - M\alpha) : 1 \le j \le s, 1 \le p \le N, \alpha \in \mathbb{Z}^d\}$ such that for any $f \in V_\varphi^2$*

$$f = m \sum_{j=1}^s \sum_{p=1}^N \sum_{\alpha \in \mathbb{Z}^d} \mathcal{L}_j f^{(p)}(M\alpha) S_j^p(t - M\alpha) \quad in \, L^2(\mathbb{R}^d)^{(r)}. \quad (1.7)$$

**Proof.** Part $(c)$ in Lemma 1.2 proves that conditions $(a)$ and $(b)$ are equivalent.

If $A_G > 0$ then for any $1 \le p \le N$, $\operatorname{ess\,inf}_{x \in [0, 1/N)^d} \det\left[G_p^*(x)G_p(x)\right] > 0$ and consequently, there exists the pseudo-inverse matrix

$$G_p^\dagger(x) = \left[G_p^*(x)G_p(x)\right]^{-1} G_p^*(x).$$

Moreover, its entries are essentially bounded and its first row satisfies (1.6). Therefore, $(a)$ implies $(c)$.

Next, we will prove that the condition $(c)$ implies $(d)$. Since we have assumed that $g_{j,p}(x) \in L^\infty\left([(p-1)/N, p/N)^d\right)$ for any $1 \le j \le s$ and $1 \le p \le N$, Lemma 1.2(b) proves that

$$\left\{\overline{g_{j,p}(x)}e_\alpha(M^T x) : 1 \le j \le s, 1 \le p \le N, \alpha \in \mathbb{Z}^d\right\}$$

is a Bessel sequence in $L^2[0, 1)^d$. The same argument proves that

$$\left\{m d_j^p(x) e_\alpha(M^T x) : 1 \le j \le s, 1 \le p \le N, \alpha \in \mathbb{Z}^d\right\}$$

is also a Bessel sequence in $L^2[0,1]^d$. By (1.4) and (1.6), these two Bessel sequences satisfy

$$F(x) = m \sum_{j=1}^{s}\sum_{p=1}^{N}\sum_{\alpha\in\mathbb{Z}^d} \left\langle F(\cdot), \overline{g_{j,p}(\cdot)}e_\alpha(M^T\cdot) \right\rangle d_j^p(x)e_\alpha(M^Tx), F \in L^2[0,1]^d.$$

Hence, they form a pair of dual frames for $L^2[0,1]^d$ (see [7, Lemma 5.6.2]). Since $S_j^p(t - M\alpha) = T_\varphi[d_j^p(\cdot)e_\alpha(M^T\cdot)](t)$ and $T_\varphi$ is an isomorphism, the sequence $\{S_j^p(t - M\alpha) : 1 \leq j \leq s, 1 \leq p \leq N, \alpha \in \mathbb{Z}^d\}$ is a frame for $V_\varphi^2$.

Last, we prove that the condition (d) implies (b). Notice that since we have assumed that $\{\overline{g_{j,p}(x)}e_\alpha(M^Tx) : 1 \leq j \leq s, 1 \leq p \leq N, \alpha \in \mathbb{Z}^d\}$ is a Bessel sequence with bound $B_G/m$ and

$$(\mathcal{L}_j f^{(p)})(M\beta) = \left\langle F(\cdot), \overline{g_{j,p}(\cdot)}e_\beta(M^T\cdot) \right\rangle_{L^2\left([(p-1)/N, p/N)^d\right)}.$$

For each $f \in V_\varphi^2$, we have

$$\sum_{p=1}^{N}\sum_{j=1}^{s}\sum_{\alpha\in\mathbb{Z}^d} \left|\mathcal{L}_j f^{(p)}(M\alpha)\right|^2 \leq \frac{B_G}{m}\|F\|_{L^2[0,1]^d}^2 \leq \frac{B_G\|T_\varphi^{-1}\|_{oper}}{m}\|f\|_{L^2(\mathbb{R}^d)^{(r)}}^2.$$

If $\{S_j^p(t - M\alpha) : 1 \leq j \leq s, 1 \leq p \leq N, \alpha \in \mathbb{Z}^d\}$ is a frame for $V_\varphi^2$, then the formula (1.7) gives

$$\begin{aligned}
\|f\|_{L^2(\mathbb{R}^d)^{(r)}}^2 &= m^2 \left\|\sum_{j=1}^{s}\sum_{p=1}^{N}\sum_{\alpha\in\mathbb{Z}^d} \mathcal{L}_j f^{(p)}(M\alpha)S_j^p(t - M\alpha)\right\|_{L^2(\mathbb{R}^d)^{(r)}}^2 \\
&\leq m^2 C \sum_{p=1}^{N}\sum_{j=1}^{s}\sum_{\alpha\in\mathbb{Z}^d} \left|\mathcal{L}_j f^{(p)}(M\alpha)\right|^2,
\end{aligned}$$

where $C$ is a Bessel bound for $\{S_j^p(t - M\alpha) : 1 \leq j \leq s, 1 \leq p \leq N, \alpha \in \mathbb{Z}^d\}$. Hence, the set $\{\mathcal{L}_1, \mathcal{L}_2, \cdots, \mathcal{L}_s\}$ is an $M$-stable filtering sampler for $V_\varphi^2$. $\square$

# References

[1] X. Zhou and W. Sun, On the sampling theorem for wavelet subspaces, J. Fourier Anal. Appl. **5** (1999), 347–354.

[2] Z. Shang and W. Sun and X. Zhou, Vector sampling expansions in shift invariant subspaces, J. Math. Anal. Appl. **325** (2007), 898–919.

[3] A.G. García and G. Pérez-Villalón, Multivariate generalized sampling in shift-invariant spaces and its approximation properties, J. Math. Anal. Appl. **355** (2009), 397–413.

[4] A.G. García and G. Pérez-Villalón, G., Dual frames in L2 (0, 1) connected with generalized sampling in shift-invariant spaces, Appl. Comput. Harmon. Anal **20** (2006), 422–433.

[5] A. Aldroubi, Q. Sun, W.-S. Tang, Convolution, average sampling, and a Calderon resolution of the identity for shift-invariant spaces, J. Fourier Anal. Appl **11** (2005), 215–244.

[6] P. Wojtaszczyk, "A Mathematical Introduction to Wavelets," Cambridge University Press, Cambridge, 1997.

[7] O. Christense, "A Introduction to Frames and Riesz Bases," Birkhäuser, Boston, 2003.